\documentclass{aa}  

\usepackage{amsmath,amssymb}
\usepackage{cancel}
\usepackage{graphicx}
\DeclareGraphicsExtensions{.png,.pdf}
\usepackage{txfonts}
\usepackage[english]{babel}
\usepackage[utf8]{inputenc}
\usepackage[T1]{fontenc}
\usepackage{fontenc}
\usepackage{soul}
\allowdisplaybreaks

\newcommand\Sect{Sect.\ }

\newcommand\Eq{Eq.\ }
\newcommand\Eqs{Eqs.\ }
\newcommand\Fig{Fig.\ }

\newcommand\primed{^{\prime}}
\newcommand\primedprimed{^{\prime\prime}}

\newcommand{\beq}{\begin{equation}}
\newcommand{\eeq}{\end{equation}}

\newcommand\omegabar{\bar{\omega}}
\newcommand\const{\,{\rm const}}

\newcommand\cs{c_{\rm s}}
\newcommand\csiso{c_{\rm s,iso}}

\newcommand\sone{s_1}
\newcommand\kone{k_1}
\newcommand\ktwo{k_2}
\newcommand\sonetwo{s_{1,2}}
\newcommand\stwo{s_2}

\newcommand\dd{{\rm d}}

\newcommand\de{\partial}
\newcommand\Deltaomega{\Delta_{\omega}}
\newcommand\Deltas{\Delta_{s}}

\newcommand\deltarho{\delta \rho}

\newcommand\grad{{\nabla}}
\newcommand{\gv}{{\boldsymbol g}}

\newcommand{\imaginary}{{\rm i}}
\newcommand{\im}{\imaginary}

\newcommand\kx{k_x}

\newcommand\kz{k_z}

\newcommand{\kJ}{k_{\rm J}}
\newcommand{\kstar}{k_*}
\newcommand\kv{{\boldsymbol k}}

\newcommand\Lcal{\mathcal{L}}

\newcommand\Phihat{\hat{\Phi}}

\newcommand\Phiext{\Phi_{\rm ext}}
\newcommand\Phitot{\Phi_{\rm tot}}

\newcommand\press{p}

\newcommand\presshat{\hat{\press}}

\newcommand{\qhat}{\hat{q}}

\newcommand\rhohat{\hat{\rho}}

\newcommand{\uv}{{\boldsymbol u}}
\newcommand{\uvhat}{\hat{\boldsymbol u}}

\newcommand{\ux}{u_x}
\newcommand{\uxhat}{\hat{u}_x}

\newcommand{\uz}{u_z}
\newcommand{\uzhat}{\hat{u}_z}

\usepackage[]{hyperref}
\hypersetup{colorlinks=true,linkcolor=red,citecolor=blue,urlcolor=magenta}

\defcitealias{Nip23}{N23}

\begin{document}

\titlerunning{Jeans criterion for hydrostatic and infalling gas}
\title{On the Jeans criterion for hydrostatic and infalling gas}

\authorrunning{C.\ Nipoti}

\author{Carlo Nipoti}

\institute{
 Dipartimento di Fisica e Astronomia ``Augusto Righi'', Alma Mater Studiorum - Università di Bologna, via Gobetti 93/2, 40129, Bologna, Italy\\
\email{carlo.nipoti@unibo.it}\\
}

\date{Accepted, January 10, 2026}

\abstract{}{We study the local gravitational instability of
  non-rotating astrophysical fluids allowing for the presence of an
  external gravitational potential in addition to the fluid
  self-gravity.}{We present a self-consistent linear-perturbation
  analysis taking into account pressure and density gradients in the
  background medium. We explore two different steady-state
  configurations for the unperturbed gas: hydrostatic equilibrium and
  infall into a gravitational potential well.}{We show that in both
  cases the instability criterion is the classical Jeans criterion,
  which, contrary to previous claims, is not modified by the presence
  of the external gravitational field. {While in the case of
    hydrostatic equilibrium linear local perturbations are always
    gravitationally stable, the conditions for gravitational
    instability can be met in the case of infalling gas, also in the
    presence of additional non-gravitational forces such as that due
    to a wind.}}{{We conclude that the Jeans criterion can have a
    role in regulating the formation of clumps and star clusters in
    streams or shells of gas infalling into galactic gravitational
    potential wells, as well as, on smaller scales, the fragmentation
    of gas in collapsing molecular clouds.}}


   \maketitle

 \section{Introduction}
 \label{sec:intro}

The local gravitational instability of astrophysical gas is a
fundamental mechanism, which is believed to have a role, for instance,
in the formation of gas clumps and star clusters in galaxies
(e.g.\ section 8.3 of \citealt{CFN19}), in the fragmentation of
molecular clouds \citep{Fie08,Vaz25}, and in the formation of planets
\citep{Kra16}.  {The conditions for local gravitational instability are
more easily found in rotation-supported gaseous discs
\citep{Too64,Gol65} than in pressure-supported gaseous systems, where
local self-gravitating perturbations are stabilized by the thermal or
turbulent pressure (\citealt{Bin08}, \citealt{Ber14}; see also
\citealt{Nip23}).}

Recent studies on the formation of gas clumps and stellar clusters in
galaxies have pointed out that there are two channels in which the gas
can fragment via local gravitational instability: not only in rotating
gaseous discs, but also in low-angular momentum gas infalling in the
galaxy gravitational potential either in filamentary streams
\citep{Fre14,Man18,van23} or in roughly spherical shells
\citep{Dek23}.  {In this paper we focus on the case of infalling gas
and present a linear-perturbation analysis of gas steadily inflowing in
the presence of an external gravitational potential (generated, for
instance, by the galaxy dark-matter halo and stellar components) in
addition to the fluid self-gravity.}  We address the problem
self-consistently, taking into account that the infalling gas can have
density, pressure and velocity gradients, as it is natural in the
presence of a gravitational field. We also discuss the effect of
additional non-gravitational forces such as that due to a wind.

Before addressing the case of infalling gas, we find it convenient to
present a self-consistent local gravitational stability analysis of a
stratified gas in hydrostatic equilibrium in its own gravity plus an
external gravitational potential. Our analysis of the hydrostatic case
can be seen as a revisited \citet{Jea02} gravitational-stability
analysis, in which, by allowing for the presence of gradients in the
background quantities, we do not resort to the so-called Jeans swindle
\citep[e.g.][]{Bin08}.

The paper is organized as follows. \Sect\ref{sec:refeq} introduces the
relevant equations. \Sect\ref{sec:hydrostatic} and \ref{sec:infalling}
present, respectively, the analyses of the hydrostatic and infalling
gas. Our results are compared with previous work in
\Sect\ref{sec:comparison}. \Sect\ref{sec:concl} concludes.

  \section{Reference equations}
  \label{sec:refeq}

We want to study the conditions for local gravitational instability in
gas either in hydrostatic equilibrium or infalling into a gravitational
potential. {For an adiabatic inviscid unmagnetized fluid, the relevant set of
equations is
  \begin{equation}
    \begin{split}
&{\de \rhohat \over \de t}+\grad\cdot(\rhohat\uvhat)=0,\\
&{\partial \uvhat \over \partial
t}+\left(\uvhat\cdot\grad\right) \uvhat =-\frac{1}{\rhohat}\grad\presshat -\grad\Phihat-\grad\Phiext,\\
      &\left({\partial \over \partial t}+\uvhat\cdot\grad\right) \ln \left(\frac{\presshat}{ \rhohat^{\gamma}}\right)=0,\\
      &\nabla^2\Phihat=4\pi G\rhohat,
    \end{split}
    \label{eq:vectsyst}
  \end{equation}
where the first three equations are the hydrodynamic mass, momentum
and entropy equations, and the last equation is the Poisson equation.
 Here $\rhohat$, $\presshat$, $\uvhat$ and $\Phihat$ are the gas density, pressure,
velocity and gravitational potential, respectively. $\Phiext$ is a
fixed external gravitational potential and $\gamma$ is the adiabatic
exponent.}

\begin{figure}
  \centerline{\includegraphics[width=0.4\textwidth]{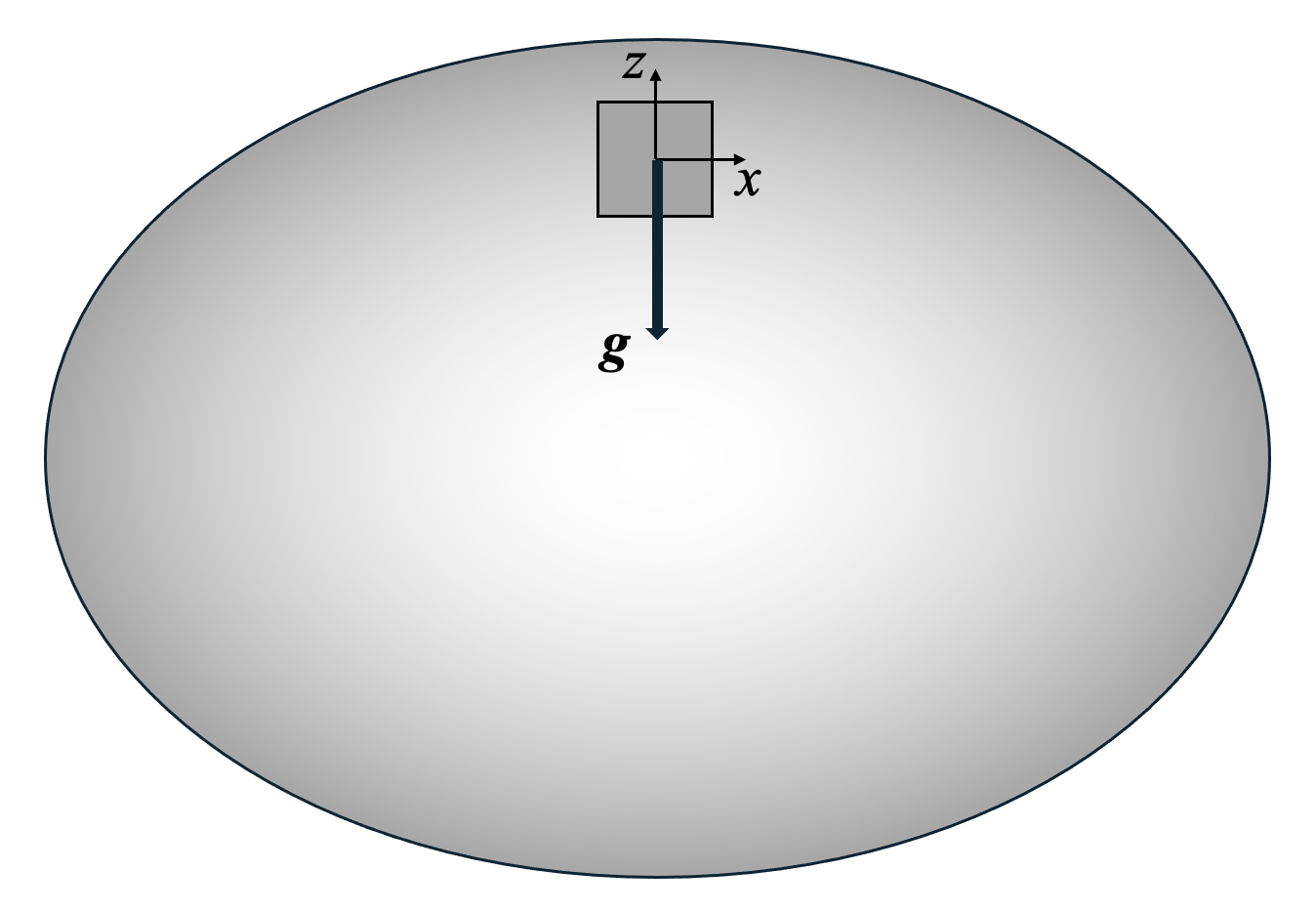}}
  \caption{Scheme of the Cartesian $x$-$z$ coordinate system adopted
    to describe locally a parcel of gas immersed in a gravitational
    potential. $\gv$ is the gravitational field produced by the
    unperturbed total gravitational potential at the position of the
    parcel of gas.}
\label{fig:cartoon}
\end{figure}

Without loss of generality, we work in a two-dimensional Cartesian
coordinate system $(x,z)$, in which the $z$ axis has the same
direction as the unperturbed gravitational field, but opposite
orientation (see \Fig\ref{fig:cartoon}).  For instance, in the case of
a spherical gravitational potential $z$ is the radial direction and
$x$ is a generic tangential direction. Given that we perform a local
perturbation analysis, the curvature is negligible, so for our purpose
Cartesian coordinates work as well as curvilinear coordinates.
{In our system of coordinates the system of equations
(\ref{eq:vectsyst}) becomes
\begin{equation}
\begin{split}
&{\de \rhohat \over \de t}+\frac{\de \left(\rhohat\uxhat\right)}{\de x}+\frac{\de\left( \rhohat\uzhat\right)}{\de z}=0,\\
&\frac{\de \uxhat}{\de t}+\uxhat\frac{\de \uxhat}{\de x}+\uzhat\frac{\de \uxhat}{\de z}=-\frac{1}{\rhohat}\frac{\de \presshat}{\de x} -\frac{\de \Phihat}{\de x}-\frac{\de \Phiext}{\de x},\\
&\frac{\de \uzhat}{\de t}+\ux\frac{\de \uzhat}{\de x}+\uz\frac{\de \uzhat}{\de z}=-\frac{1}{\rhohat}\frac{\de \presshat}{\de z} -\frac{\de \Phihat}{\de z}-\frac{\de \Phiext}{\de z},\\
&\left(\frac{\de}{\de t}+\uxhat\frac{\de }{\de x}+\uzhat\frac{\de }{\de z}\right) \ln \left(\frac{\presshat}{ \rhohat^{\gamma}}\right)=0,\\
&\frac{\de ^2\Phihat}{\de x^2}+\frac{\de ^2\Phihat}{\de z^2}=4\pi G\rhohat.
\end{split}
    \label{eq:coordsyst}
\end{equation}

Let us consider a generic quantity $\qhat=\qhat(x,z,t)$ describing a
property of the fluid (such as $\rhohat$, $\presshat$, $\Phihat$ or
any component of $\uvhat$): $q$ can be written as $\qhat=q+\delta q$,
where the (time independent) quantity $q$ describes the stationary
unperturbed fluid and the (time dependent) quantity $\delta q$
describes the Eulerian perturbation. The external gravitational
potential $\Phiext$ is assumed to be independent of time and not affected
by the perturbation.}
  
Though we are mainly interested in the case of gas infalling into a
gravitational potential, it is instructive to discuss first the case
in which the unperturbed gas is in hydrostatic equilibrium.

  \section{Hydrostatic gas}
\label{sec:hydrostatic}
  
\subsection{Unperturbed system}

Under the assumption of hydrostatic equilibrium, the unperturbed
configuration is a static solution of the system of equations
(\ref{eq:coordsyst}), so all the time derivatives and all the
components of $\uv$ are null. We further assume that the quantities
describing the unperturbed system can have non-null gradients only
along the $z$ direction, so $\de q/\de x=0$.  In the following we use
$\primed$ to indicate derivatives w.r.t.\ $z$, so $q\primed\equiv\dd
q/\dd z$ and $q\primedprimed\equiv\dd^2 q/\dd z^2$ and for any
unperturbed quantity $q=q(z)$.

Under these assumptions the unperturbed medium is in equilibrium in
the total gravitational potential and thus satisfies
$\press\primed/\rho= -\Phitot\primed$, where $\Phitot=\Phi+\Phiext$,
and $\Phi\primedprimed=4\pi G\rho$.  Given that $\uv=0$, the
stationary entropy equation does not put constraints on the entropy
profile, so the unperturbed system can have $\sigma\primed\neq 0$
where $\sigma\equiv \ln \left({\press}{ \rho^{-\gamma}}\right)$ is the
normalized specific entropy.

\subsection{Perturbations and linearized equations}
\label{sec:static_pert}

{We take small ($|\delta q|\ll|q|$) plane-wave perturbations with wavevector $\kv=(\kx,\kz)$
in the form $\delta q \propto\exp{[\imaginary(\kx x +\kz z -\omega
    t)]}$, where $\omega$ is the frequency.}  We further assume that
the perturbations are local, so $|\kz|\gg 1/\ell$, where $\ell\equiv
|q/q\primed|$ is the characteristic length over which any quantity $q$
varies in the unperturbed configuration at the position of the
disturbance.  Under these assumptions, the linearized perturbed
equations are
\begin{equation}
\begin{split}
&-\im \omega \delta\rho+\im \kx \rho \delta\ux+\im \kz \rho \delta\uz=0,\\
  &-\im \omega\delta\ux=-\im \frac{\kx}{\rho}\delta\press-\im\kx\delta\Phi,\\
  & -\im \omega\delta\uz=-\im \frac{\kz}{\rho}\delta\press+\frac{\press\primed}{\rho^2}\delta\rho-\im\kz\delta\Phi,\\
  &-\im \omega \frac{\delta \press}{\press}+\im\gamma\omega\frac{\delta\rho}{\rho}+\sigma\primed\delta\uz=0,\\
&-k^2\delta\Phi=4\pi G\deltarho,  
\end{split}
\label{eq:lin_sys_stat}
 \end{equation}
where $k^2=\kx^2+\kz^2$. In the first equation of the above system we
omitted the term $\rho\primed\delta\uz$, which is negligible compared
with $\im \kz \rho \delta\uz$ for local perturbations.  The system
(\ref{eq:lin_sys_stat}) reduces to the biquadratic dispersion relation
\begin{equation}
\omega^4+(4\pi G\rho-\cs^2k^2)\omega^2+\cs^2\kx^2N^2=0,
\label{eq:disprelstat}
\end{equation}
where $\cs^2\equiv\gamma\press/\rho$ is the sound speed squared and
$N^2\equiv-\sigma\primed\press\primed/(\gamma\rho)$, which can be
null, negative or positive, is the buoyancy frequency squared.

\subsection{Wavenumbers of local perturbations}
\label{sec:wavenumber}

The dispersion relation (\ref{eq:disprelstat}) was derived assuming
that the perturbations are local, so $k$ must be larger not only than
$1/\ell$, but also than $1/\Lcal$, where $\Lcal$ is the macroscopic
length scale of the gaseous system \citep[see][]{Nip23}.  Before
proceeding to the analysis of the dispersion relation, it is useful to
compare the wavenumbers of such local perturbations with two
characteristic wavenumbers associated with the physical properties of
the unperturbed system. The first is the Jeans wavenumber
\begin{equation}
\kJ=\frac{\sqrt{4\pi G\rho}}{\cs};
\label{eq:kj}
\end{equation}  
the second, defined only when $N^2>0$, is
\begin{equation}
\kstar\equiv\frac{\sqrt{N^2}}{\cs},
\end{equation}
that is the wavenumber of a disturbance for which the sound-wave
frequency is the same as the buoyancy frequency. We now show that, for
the hydrostatic case here considered, local perturbations have $k$
larger than both $\kJ$ and $\kstar$.

As it is well known \citep[e.g.][]{Bin08,Ber14}, for a
self-gravitating cloud in hydrostatic equilibrium $\kJ\sim 1/\Lcal$. A
similar result is obtained for gas in hydrostatic equilibrium in the
presence of an external gravitational potential in addition to its own
gravitational field.  Let us assume that the total gravitational
potential is generated by a mass $fM$, where $M$ is the gas mass and
$f>1$ accounts for the mass generating the external potential.  In
this total potential, a hydrostatic gas with sound speed $\cs$ has
characteristic size $\Lcal \approx G f M / \cs^2 $.  Using $M\approx
4\pi \rho \Lcal^3/3$, where $\rho$ is the mean gas density, we get
$\Lcal^2\approx 3 \cs^2/(4\pi f G \rho)\approx 3/(f\kJ^2)$.  So, for a
gaseous system in equilibrium with its self-gravity and an external
potential, $\kJ$ is of the order of or smaller than $1/\Lcal$. Thus,
local perturbations have $k$ larger than $\kJ$.

From the definition of $N^2$ (\Sect\ref{sec:static_pert}), assuming
$\press\primed<0$, as it is usual, and $\sigma\primed>0$, in order to
have $N^2>0$, we get
\begin{equation}
\kstar^2=\frac{N^2}{\cs^2}=\frac{|\press\primed|\sigma\primed}{\gamma
  \rho\cs^2}
=\frac{|\press\primed|}{\press}\frac{\sigma\primed}{\gamma^2} \sim
\frac{1}{\ell^2},
\label{eq:kstar}
\end{equation}
given that $\press\primed/\press\sim 1/\ell$ and
$\sigma\primed\sim 1/\ell$. Thus $\kstar\sim 1/\ell$ and local
perturbations have $k>\kstar$.

\subsection{Stability criterion}
\label{sec:static_crit}

In Appendix \ref{sec:disprel} we report the full formal analysis of
the dispersion relation (\ref{eq:disprelstat}), considering all
wavevectors. Combining the results of Appendix \ref{sec:disprel} with
the physical restrictions on the perturbation wavenumber discussed in
\Sect\ref{sec:wavenumber}, we draw the following conclusions about
stability or instability of self-gravitating perturbations.
\begin{itemize}
\item When $N^2=0$ the condition for stability is $k \geq\kJ$, so all
  local perturbations are stable. We cannot draw conclusions about
  perturbations with $k\lesssim \kJ$, because these perturbations are
  not local.

\item When $N^2<0$ all local perturbations are unstable, but the
  instability is essentially convective (e.g.\ \citealt{Tas78}), and
  thus not expected to lead to clump formation. Mathematically, the
  convective nature of the instability is apparent if one considers
  that the fastest-growing unstable mode has growth rate
  $\left|\omega^2_1\right|^{1/2}$, with $\omega^2_1$ given by
  \Eq(\ref{eq:omegaonetwo}). In the relevant regime $k>\kJ$ the
  absolute value of $\omega^2_1$ can be rewritten as
  \begin{equation}
\left|\omega^2_1\right|=\frac{\cs^2}{\sqrt{2}}\left(k^2-\kJ^2\right)\left[\sqrt{1+\frac{4\kx^2|N^2|}{\cs^2(k^2-\kJ^2)^2}}-1\right].
  \end{equation}  
When $k/\kJ$ is sufficiently large, we can replace $k^2-\kJ^2$ with $k^2$
and then, using $|N^2|\ll\cs^2k^2$ (see \Eq\ref{eq:kstar}), we can 
Taylor-expand the square root and get
 \begin{equation}
\left|\omega^2_1\right|\approx \frac{\kx^2}{k^2}\left|N^2\right|,
 \end{equation}
which is the convective-instability growth rate squared.
 
\item When $N^2>0$, limiting ourselves to the relevant wavenumbers with $k>\kJ$
  and $k>\kstar$, we never have monotonic instability. Perturbations
  with $k>\ktwo$ (where $\ktwo\geq \kJ$ is a critical wavenumber
  defined in Appendix \ref{sec:disprel}) are stable. Formally there is
  room for overstable disturbances with $\kJ<k<\ktwo$, but
\begin{equation}
  \ktwo^2<2\left(\kJ^2+2\kstar^2\right)\lesssim 2\left(\frac{1}{\Lcal^2}+2\frac{1}{\ell^2}\right),
\end{equation}  
so $k<\ktwo$ perturbation cannot be considered local. We conclude that
all local perturbations are stable.

\end{itemize}

In conclusions, local gravitational instability never occurs in a gas
that is in hydrostatic equilibrium, either in the presence or in the
absence of an external gravitational potential.

\section{Infalling gas}
\label{sec:infalling}

\subsection{Unperturbed system}
\label{sec:unpert_infall}

Also when the gas is infalling, the unperturbed configuration is
assumed to be a stationary solution of the system of equations
(\ref{eq:coordsyst}), so all time derivatives are null. Given that the
fluid is inflowing, we have $\uz< 0$, while $\ux=0$.  We further
assume that the quantities describing the unperturbed system can have
non-null gradients only along the $z$ direction, so $\de q/\de x=0$ for
any unperturbed quantity $q$. Under these assumptions the unperturbed
medium is a stationary Bondi-like solution \citep{Bon52} with
$\rho\uz=\const$, $\press\rho^{-\gamma}=\const$ (thus
$\sigma\primed=0$), $\uz\uz\primed=-\press\primed/\rho-\Phitot\primed$
and $\Phi\primedprimed=4\pi G\rho$.

\subsection{Perturbations and linearized equations}

Applying the same kind of plane-wave perturbations as in
\Sect\ref{sec:static_pert} and linearizing the system of equations
(\ref{eq:coordsyst}), in the case of infalling gas we get
\begin{equation}
\begin{split}
&-\im \omega \delta\rho+\im \kx \rho \delta\ux+\im \kz \rho \delta\uz+\im \kz\uz\delta\rho=0,\\
  &-\im \omega\delta\ux+\im \kz\uz\delta\ux=-\im \frac{\kx}{\rho}\delta\press-\im\kx\delta\Phi,\\
  & -\im \omega\delta\uz+\im \kz\uz\delta\uz=-\im \frac{\kz}{\rho}\delta\press+\frac{\press\primed}{\rho^2}\delta\rho-\im\kz\delta\Phi,\\
  &-\im \omega \frac{\delta \press}{\press}+\im\gamma\omega\frac{\delta\rho}{\rho}+\im \kz\uz \frac{\delta \press}{\press}-\im\gamma\kz\uz\frac{\delta\rho}{\rho}=0,\\
&-k^2\delta\Phi=4\pi G\deltarho.
\end{split}
\end{equation}
This linear system leads to the dispersion relation
\begin{equation}
  \omegabar^2=\cs^2k^2-4\pi G\rho,
  \label{eq:doppler}
\end{equation}  
where $\omegabar\equiv\omega-\kz\uz$ is the ``Doppler-shifted''
frequency \citep{Mal87}.

\subsection{Stability criterion}

Noting that the imaginary part of $\omega$ is equal to the imaginary
part of $\omegabar$, the condition for instability is $\omegabar^2<0$,
i.e. $\cs^2k^2<4\pi G \rho$, which is just the Jeans criterion
$k<\kJ$. Thus, we can have unstable perturbations, provided that the
Jeans length is small compared to the macroscopic scales of the system
(i.e. $\kJ>|\rho\primed/\rho|$, $\kJ>|\press\primed/\press|$ and
$\kJ>|\uz\primed/\uz|$). While local perturbations have $k>\kJ$ when
the unperturbed medium is in hydrostatic equilibrium
(\Sect\ref{sec:wavenumber}), there can be local perturbations with
$k<\kJ$ when the gas is infalling. This can be seen by considering,
for instance, that $\kJ$ (\Eq\ref{eq:kj}), at given $\rho$, increases
for decreasing $\cs$, while the macroscopic scale of the system
(e.g.\ $|\rho/\rho\primed|$) is not directly related to $\cs$,
because the gas is not in hydrostatic equilibrium.

{We conclude that the local gravitational stability or instability of gas
infalling into a gravitational potential well is regulated by the
Jeans criterion. If the Jeans length is smaller than the macroscopic
scales of the system, perturbations with $k<\kJ$ are expected to be
gravitationally unstable.}

\subsection{Additional non-gravitational external forces}

The above results on infalling gas are essentially unaltered if we
allow for the presence of another external force, in addition to the
external gravitational field, such as that due to a wind directed
along $z$ and opposite to the gravitational field, in a configuration
similar to the one envisaged by \citet{Dek23} in their shell scenario
for star cluster formation.  If the wind partially balances the
gravitational field, the unperturbed gas is undergoing an infall,
which can be stationary, as described by the equations reported in
\Sect\ref{sec:unpert_infall}. In principle, if the wind, combined with
possible pressure gradients, exactly balances the gravitational force,
the unperturbed gas can be also static. In neither case the
restrictions to the perturbation wavenumber described in
\Sect\ref{sec:wavenumber} apply, because the characteristic
macroscopic scales are not determined by the equilibrium between
gravity and pressure gradients. {It follows that, also in the presence
of non-gravitational external forces, such as that exerted by a wind,
the linear stability criterion is the classical Jeans criterion, and the
conditions for instability can be met.}

\section{The role of the external gravitational potential}
\label{sec:comparison}

In some respect, the analysis presented in this paper is comparable
with that of \citet[][]{Jog13}, who studied the Jeans instability of a
static fluid in the presence of an external gravitational potential
$\Phiext$, but (different from what is done here) assuming that the
unperturbed medium is homogeneous and with no pressure gradients (see
also \citealt{Zav23}).

\citet{Jog13} found the dispersion relation $\omega^2=\cs^2 k^2-4\pi G
\rho-\Phiext\primedprimed$, to be compared with our dispersion
relation (\ref{eq:disprelstat}) that, for $N^2=0$ (as it is the case
for medium with no gradients), reduces to the classical Jeans
dispersion relation
\begin{equation}
\omega^2=\cs^2 k^2-4\pi G\rho.
\label{eq:jeans_disp_rel}
\end{equation}
$\Phiext$ does not appear in our dispersion relation, because, apart
from $N^2$, all the terms containing
$\press\primed=-\rho(\Phi\primed+\Phiext\primed)$ are
self-consistently neglected during the analysis when compared with
similar terms containing $\kz\press$ instead of $\press\primed$.  As far
as we can tell, the whole analysis of \citet{Jog13} is undermined by
an ill-defined unperturbed solution, in which a static, homogeneous
gas with no pressure gradients is assumed to be in a steady-state in
the presence of a non-vanishing external gravitational field.  More
specifically, using the notation of the present paper and defining the
isothermal sound speed $\csiso=\gamma^{-1/2}\cs$, in equation 8 of
\citet{Jog13} the term $\delta\rho\grad\Phiext$ is negligible compared
to the term $\grad(\csiso^2\delta\rho)$, because
\begin{equation}
\delta\rho\grad\Phiext=-\delta\rho\frac{\grad\press}{\rho}
=-\csiso^2 \delta\rho\frac{\grad\press}{\press}\sim\frac{1}{\ell} \csiso^2 \delta\rho,
\end{equation}
where $\grad\press$ would be the pressure gradient required to balance $\grad\Phiext$ in the unperturbed configuration, while
\begin{equation}
\grad(\csiso^2\delta\rho)=\csiso^2\delta\rho \kv+\csiso^2\delta\rho\frac{\grad\csiso^2}{\csiso^2}\sim k  \csiso^2\delta\rho,
\end{equation}
where we have used $k\gg |\grad\csiso^2|/\csiso^2\sim 1/\ell$. 

{Another paper presenting an analysis related to this work is
  \citet[][see also \citealt{Col20} and \citealt{Dum25}]{Lee18}, who
  studied the stability of non-linear density fluctuations in the
  presence of an external field. \citet{Lee18} concluded that the
  tidal forces due to the external field can have a stabilizing
  effect. This result is not in tension with our findings, because,
  while we focused on linear and local disturbances, the perturbations
  considered by \citet{Lee18} not only are non-linear (with typical
  density contrast $\gtrsim 1$), but are also non-local (i.e.\ such
  that the external gravitational field has non-negligible gradient
  across the size of the perturbation). Whatever the behaviour of such
  non-linear and non-local disturbances, an underlying question is how
  such disturbances are produced in the first place. The results of
  the present work can help understand under which conditions they can
  originate from small local perturbations.}

{In conclusion, even in the presence of an external gravitational
  field, when the unperturbed medium is in hydrostatic equilibrium
  with no entropy gradients ($\sigma\primed=0$), the dispersion
  relation for linear local self-gravitating perturbations is the
  classical Jeans dispersion relation (\ref{eq:jeans_disp_rel}). When
  $\sigma\primed\neq 0$ the dispersion relation takes the more general
  form of \Eq(\ref{eq:disprelstat}).} In neither case the stability
criterion is modified by the presence of the external
field\footnote{Though $N^2$ can also be written as
$\sigma\primed\Phitot\primed/(\gamma\rho^2)$, with
$\Phitot=\Phi+\Phiext$, $N^2$ enters the stability criterion only with
its sign, which is determined only by the sign of $\sigma\primed$,
because $\Phitot\primed>0$ by construction.}.

For essentially the same reasons, the external field does not appear
in the dispersion relation when the unperturbed gas is infalling
(\Eq\ref{eq:doppler}). This implies, for instance, that the classical
Jeans criterion can be applied to study the stability of infalling
gaseous streams such as those studied by \citet{Man18}.

\section{Conclusions}
\label{sec:concl}

We have revisited the problem of the local gravitational instability
of non-rotating astrophysical gas, considering the cases of
hydrostatic gas and of gas infalling into a gravitational potential
well. {We have shown that in both cases the linear gravitational stability
criterion is the classical Jeans criterion, which is not modified by
the presence of an external gravitational potential.}

While all local perturbations turn out to be stable in the hydrostatic
case, the conditions for local instability can be met in the case of
infalling gas, even in the presence of additional non-gravitational
forces, such as that produced by a wind. {Taken at face value, our
  results suggest that the classical Jeans criterion can regulate the
  formation of gas clumps and star clusters in gas infalling into
  galactic gravitational potentials, as well as, on smaller scales,
  the fragmentation of gas in collapsing molecular clouds.  However, a
  more realistic description of these processes should include also
  other ingredients, such as magnetic fields, turbulence, radiative
  cooling and a more complex geometry \citep[e.g.][]{Hos17}, that were
  neglected in our analysis and that can have an important role.  }

\begin{acknowledgements} 
{I am grateful to Alessandro Romeo and to an anonymous referee for
  helpful comments.}  The research activities described in this paper
have been co-funded by the European Union – NextGenerationEU within
PRIN 2022 project n.20229YBSAN - Globular clusters in cosmological
simulations and in lensed fields: from their birth to the present
epoch.
\end{acknowledgements}

\bibliographystyle{aa}
\bibliography{biblio_igi.bib}

\appendix

\section{Analysis of the dispersion relation (\ref{eq:disprelstat})}
\label{sec:disprel}

The dispersion relation (\ref{eq:disprelstat}) can be written
in the form
\begin{equation}
  \omega^4
  +(A-s)\omega^2
  +\xi C s=0,  
\label{eq:disprelstat_norm}
\end{equation}
where $A\equiv4\pi G\rho>0$, $s\equiv\cs^2k^2>0$, $\xi\equiv \kx^2/k^2$, by definition such that $0\leq \xi \leq 1$, and $C\equiv N^2$, which can be null, negative or positive. We now examine each of these three cases. 

\begin{enumerate}

\item \underline{The case $C=0$}.
When $C=0$ the dispersion relation becomes  $\omega^2=s-A$, which implies
stability for $s\geq A$ and instability for $s<A$.

\item \underline{The case $C<0$}.
The discriminant of equation (\ref{eq:disprelstat_norm}),
\begin{equation}
\Deltaomega=(A-s)^2-4\xi C s,
\label{eq:deltaomega}
\end{equation}
is always positive when $C<0$, which implies that $\omega^2$ is real.

The roots of equation (\ref{eq:disprelstat_norm}) are
\begin{equation}
\omega^2_{1,2}=\frac{1}{2}\left(s-A\pm \sqrt{\Deltaomega}\right),
\label{eq:omegaonetwo}
\end{equation}
defined so that $\omega^2_1\leq \omega^2_2$.
We have stability when $\omega^2_1\geq 0$, i.e.\
\begin{equation}
  s-A \geq \sqrt{(A-s)^2-4\xi C s},
\label{eq:stab_s}
\end{equation}
and instability when $\omega^2_1<0$, i.e.\
\begin{equation}
  s-A<\sqrt{(A-s)^2-4 \xi C s}.  
\label{eq:instab_s}
\end{equation}
The latter condition, which is evidently satisfied when $s<A$, when
$s\geq A$ becomes $4\xi C s<0$, which is always the case when $C<0$.
We conclude that all perturbations are unstable when $C<0$.

\item \underline{The case $C>0$}.
  Expanding \Eq(\ref{eq:deltaomega}) we get
\begin{equation}
\Deltaomega(s)=s^2-2(A+2\xi C)s+A^2,
\end{equation}
which has  discriminant 
\begin{equation}
\Deltas=4(A+2\xi C)^2- 4 A^2=16 (A\xi C+\xi^2 C^2),
\label{eq:deltas}
\end{equation}
which is always positive when $C>0$. 
The zeros of $\Deltaomega(s)$ are
\begin{equation}
\sonetwo=(A+2\xi C)\pm 2\sqrt{A\xi C+\xi^2 C^2},
\end{equation}
defined so that $\sone\leq \stwo$. It is useful to note that
$0<\sone<A+2C$ and $\stwo<2A+4 C$.  When $\sone< s <\stwo$ we have
$\Deltaomega^2<0$ and thus overstability.  When either $s<\sone$ or $
s >\stwo$, $\Deltaomega^2>0$, so, using \Eqs(\ref{eq:stab_s}) and
(\ref{eq:instab_s}), we infer that perturbations with $s>A$ are
stable, while perturbations with $s<A$ are monotonically unstable.
Given that $\stwo>A$, we conclude that, when $C\geq 0$, perturbations
with $s>\stwo$ are stable and the only monotonically unstable
perturbations are those with $s<\sone$ and $s<A$.
\end{enumerate}

In summary, recalling that $k^2=s/\cs^2$ and $\kJ^2=A/\cs^2$, and defining $\kone$ and $\ktwo$ such that $\kone^2=\sone/\cs^2$ and $\ktwo^2=\stwo/\cs^2$, we have the following criteria.
\begin{itemize}

  \item When $N^2=0$ the condition for stability is $k>\kJ$.

\item When $N^2<0$ perturbations with any $k$ are unstable.

\item When $N^2>0$ we have overstability when $\kone<k<\ktwo$, stability when $k>\ktwo$, and monotonic instability when $k<\kJ$  and $k<\kone$ .

\end{itemize}

\end{document}